# Robust entanglement

H. Häffner[1,2], F. Schmidt-Kaler[1], W. Hänsel[1], C. F. Roos[1,2], T. Körber[1,2], M. Chwalla[1], M. Riebe[1], J. Benhelm[1,2], U. D. Rapol[1,2], C. Becher[1], R. Blatt[1,2]

[1]*Institut für Experimentalphysik, Universität Innsbruck,*
*Technikerstraße 25, A-6020 Innsbruck, Austria*
and
[2]*Institut für Quantenoptik und Quanteninformation,*
*Österreichische Akademie der Wissenschaften, Otto-Hittmair-Platz 1, A-6020 Innsbruck, Austria*

It is common belief among physicists that entangled states of quantum systems loose their coherence rather quickly. The reason is that any interaction with the environment which distinguishes between the entangled sub-systems collapses the quantum state. Here we investigate entangled states of two trapped $Ca^+$ ions and observe robust entanglement lasting for more than 20 seconds.

PACS numbers: 03.65 Ud, 03.67 Mn

## INTRODUCTION

"Our intuition strongly suggests that a specified entanglement, as a nonlocal property of a composed quantum system, should be very fragile under the influence of the environment" [1]. This quote exemplifies our general perception that entangled states decohere rather quickly. The reason is that any interaction with the environment which distinguishes between the entangled sub-systems collapses the quantum state [2]. Here we investigate entangled states of two trapped $Ca^+$ ions and observe robust entanglement lasting for more than 20 seconds. This observation is not only of importance for fundamental science but also for the emerging field of quantum information since entanglement is believed to be the ingredient making a quantum computer [3] much more powerful than any classical machine. Because of the fragility of entanglement physicists widely assume that it is very hard -if not impossible- to construct such a quantum computer [4]. The notion of entanglement is also strongly linked with the so-called Schrödinger-cat paradox: here absurd consequences arise when a macroscopic object (a cat) is put in a superposition of two states (alive or dead). Such situations clearly do not comply with our notion of (macroscopic) reality. As scientists, we explain this observation with the collapse of the superposition states caused by the unavoidable entanglement with the object's environment. Therefore the decoherence properties of entangled states play a central role in understanding the emergence of our classical world from quantum mechanics [2]. Consequently, there is a strong interest and need in generating entangled states and investigating their coherence properties in well controlled physical systems.

## EXPERIMENTAL SETUP, PROCEDURES AND RESULTS

For our experiments two Calcium-ions are suspended in free space inside a Paul trap with electromagnetic forces [5]. A sequence of three laser pulses addressing the ions individually creates the entangled Bell state $|\Psi\rangle = (|SD\rangle + |DS\rangle)/\sqrt{2}$. [6]. Here $|S\rangle$ and $|D\rangle$ refer to the internal states of each ion (see level scheme in Fig. 1). Via state tomography [6] we find an overlap of the experimentally generated state with the ideal one (the fidelity) of up to 96%. For this Bell state we obtain coherence times of more than 1 s, consistent with the fundamental limit set by the spontaneous decay from the $D_{5/2}$-level [6]. This observation is due to the fact that the constituents of the superposition have the same energy and are thus insensitive to fluctuations common to both ions (e.g. laser frequency and magnetic field fluctuations). Similar results have been obtained with Bell states encoded in hyperfine levels of Beryllium ions [7, 8].

As spontaneous emission ultimately limits the coherence time of entangled states encoded in the S,D - state basis one can enhance the coherence time by transferring the populations into a basis spanned by the two Zeeman sub-levels of the $S_{1/2}$ ground state: $|S_{1/2}, m_J = +1/2\rangle \equiv |0\rangle$ and $|S_{1/2}, m_J = -1/2\rangle \equiv |1\rangle$. These states do not decay spontaneously, and by choosing a superposition with equal energies, e.g. $|01\rangle + |10\rangle$, we create a "decoherence free subspace" [7]. With this approach we extend the lifetime of the entangled state by more than one order of magnitude. In particular, we coherently transfer - just after the entangling operation - the population of the $|D_{5/2}, m_J = -1/2\rangle$ state to the $|0\rangle$ state, while leaving the $|1\rangle$ population untouched (see Fig. 1). The fidelity of the resulting Bell state $|\Psi'\rangle = (|01\rangle + |10\rangle)/\sqrt{2}$ is 89 % where the loss of 7 % is due to imperfect transfer pulses. For investigating decoherence, we insert a variable delay time before analyzing the state $|\Psi'\rangle$. After a delay of 1 s, full state tomography reveals that the fidelity of the entangled state is still 86 %. We plot the respective reconstructed density matrix in Fig. 2. Since the tomographic reconstruction of the full density matrix requires many experimental cycles ($\approx 1000$), it is of advantage to employ a fidelity measure that is based on a single density matrix element and thus is easier to access. Indeed, to determine a lower bound of the fidelity, it is sufficient[9] to measure the density matrix element $\langle 01|\rho|10\rangle \equiv \rho_{01,10}$, as shown in the following: The fidelity of the entangled state $|\Psi'\rangle = (|01\rangle + e^{i\phi}|10\rangle)/\sqrt{2}$ is given by $F = \langle \Psi'|\rho|\Psi'\rangle = (\rho_{01,01} + \rho_{10,10})/2 + \Re(\rho_{01,10}e^{i\phi})$. In our implementation of the state $|\Psi'\rangle$ a magnetic field gradient leads to a deterministic evolution of the phase $\phi$ in time (see below). If we correct for this effect (i.e. determine the phase $\phi$), the fidelity becomes $F = (\rho_{01,01} + \rho_{10,10})/2 + |\rho_{01,10}|$. Using the general property of the density matrix $\sqrt{\rho_{01,01} \cdot \rho_{10,10}} \geq |\rho_{01,10}|$ and the triangular inequality we obtain the lower bound to the fidelity: $F \geq F_{\min} = 2|\rho_{01,10}|$. We plot the experimentally determined values for $F_{\min}$ in Fig. 3. and find that the fidelity is larger than 0.5 for up to 20 s; thus at least up to this time the ions were still entangled [9].

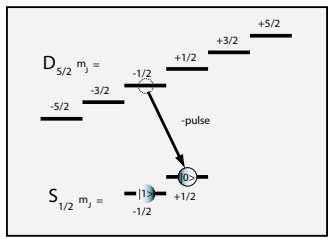

FIG. 1: Qubit levels of $^{40}$Ca$^+$: While the two-level system ($S_{1/2}$, $m_J = -1/2$ ↔ $D_{5/2}$, $m_J = -1/2$) decoheres due to spontaneous decay[5], the electron spin in the S-state ($S_{1/2}$, $m_J = -1/2$ ↔ $S_{1/2}$, $m_J = +1/2$) provides a stable quantum memory. This qubit is encoded via the S-D qubit and using a $\pi$-pulse to transfer the D-state population to the S-state manifold.

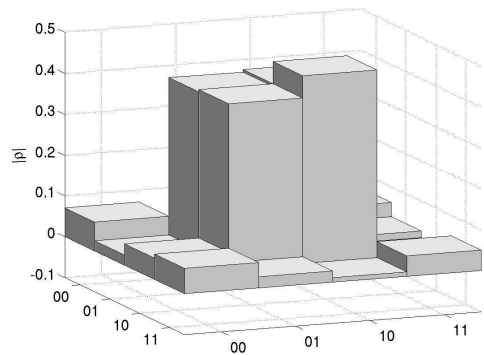

FIG. 2: Tomographically reconstructed density matrix of the Bell state $|\Psi'\rangle = (|01\rangle + |10\rangle)/\sqrt{2}$ after a waiting time of 1 s. The overlap with the ideal Bell state has dropped from 89% directly after preparation to 86% after 1 s waiting time.

We considered the following reasons for the observed decay of entanglement: slow fluctuations of the magnetic field gradient which affect the ions differently, heating of the ion crystal, residual light scattering, and collisions. The latter two were excluded experimentally: The scattering of residual light at 397 nm ($S_{1/2} - P_{1/2}$ transition) which could induce a bit flip between the states $|0\rangle$ and $|1\rangle$ was measured by preparing the ions in $|0\rangle$ or $|1\rangle$. After waiting times ranging from 2 s to 18 s we probed the state with $\pi$-pulses from the $|S_{1/2}, m_j = \pm 1/2\rangle$ state to the $|D_{5/2}, m_j = -1/2\rangle$ state. We could not detect any bit flip transitions and conclude that on average less than 1 photon is scattered in 8 minutes. The pressure in our ion trap is less than $2 \times 10^{-11}$ mbar and leads to an estimated rate of elastic collisions with the background gas of less than $3 \times 10^{-3}$ s$^{-1}$ [10]. The heating rate of the ion crystal was measured to be 1 phonon per second and vibrational mode. From the measured heating we expect a degradation of our analysis pulses which in turn would decrease the fidelity by 0.1 after 20 s. This effect, however, does not completely explain our fidelity loss. A magnetic field gradient across the ion trap lifts the energy degeneracy of the two parts of the superposition by $\Delta E = h \times 30$ Hz such that the relative phase $\phi$ of the superposition evolves as $\phi(t) = \Delta E t/\hbar$. Thus relative fluctuations of the gradient by $10^{-3}$ within the measurement time of up to 90 minutes per data point could explain the observed decay rate of $|\langle 01 | \rho | 10 \rangle|$. Generally, a slow dephasing mechanism such as fluctuations of the magnetic field gradient leads to a Gaussian decay of the coherence. A Gaussian fit to the data in Fig. 3 yields a time constant of 34(3) s for the loss of coherence of the entangled state.

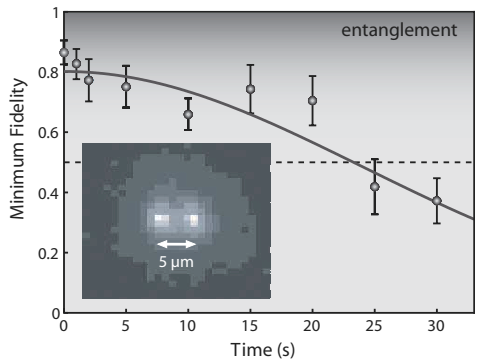

FIG. 3: Minimum fidelity of the Bell state as a function of the delay time as inferred from the density matrix element $\langle 01 | \rho | 10 \rangle$ (see text). A fidelity of more than 0.5 indicates the presence of entanglement. The inset shows the fluorescence image of two $Ca^+$ ions which were entangled in this measurement.

## DISCUSSION AND OUTLOOK

Previous experiments with single trapped $Be^+$-ions have demonstrated that single particle coherence can be kept for more than 10 minutes [11]. Here we show that also entangled states can be preserved for many seconds: the two-ion Bell states in our investigations outlive the single particle coherence time of about 1 ms in our system [12] by more than 4 orders of magnitudes. Even in the presence of an environment hostile for a single atom quantum memory, the coherence is preserved in a decoherence free subspace [7]. But more importantly, the coherence was kept non-locally, i.e. over distances which are many orders of magnitude larger than atomic or molecular dimensions. With this at hand, we envision quantum computers using these long-lived entangled states for quantum memory and for extended quantum information processing. Only twice the resources (qubits + elementary quantum gates in the decoherence free subspace) are needed to realize up to 4 orders of magnitude more operations before the quantum information is lost to the environment.

Acknowledgements: We gratefully acknowledge support by the European Commission (CONQUEST (MRTN-CT-2003-505089) and QGATES (IST-2001-38875) networks), by the ARO (No. DAAD19-03-1-0176), by the Austrian Fonds zur Förderung der wissenschaftlichen Forschung (FWF, SFB15), and by the Institut für Quanteninformation GmbH. H.H. acknowledges funding by the Marie-Curie-program of the European Union. T.K. acknowledges funding by the Lise-Meitner program of the FWF.